\documentclass{eptcs}
\providecommand{\event}{GRAPHITE 2014}
\usepackage{breakurl}

\usepackage{amsmath}
\usepackage{amssymb}
\usepackage{amsthm}
\usepackage{url}
\usepackage{stmaryrd}
\usepackage{graphicx}
\usepackage{stackrel}
\usepackage{enumitem}
\usepackage{semantic}
\usepackage{acronym}
\usepackage{caption}
\usepackage{subcaption}

\newcommand{\ignore}[1]{{}}

\newcommand{\ptrans}[0]{PTRANS}

\newtheorem{definition}{Definition}
\newtheorem{theorem}{Theorem}
\newtheorem{lemma}{Lemma}

\newcommand{\AU}[2]{\textrm{A}\ #1\ \textrm{U}\ #2}
\newcommand{\EU}[2]{\textrm{E}\ #1\ \textrm{U}\ #2}
\newcommand{\EB}[2]{\textrm{E}\ #1\ \textrm{B}\ #2}
\newcommand{\AB}[2]{\textrm{A}\ #1\ \textrm{B}\ #2}

\newcommand{\foctlmod}[4]{#1;#2;#3\vDash #4}
\newcommand{\smtmod}[2]{#1\vDash_{\textsc{FOL}}#2}

\newcommand{\satis}[1]{\textsc{Satis}(#1)}
\newcommand{\PathsBackRaw}{\textsc{Paths}_{\leftarrow}}
\newcommand{\PathsBack}[4]{\PathsBackRaw(#1,#2,#3,#4)}
\newcommand{\PathsBBackRaw}{\textsc{Paths}_{\overline{\leftarrow}}}
\newcommand{\PathsBBack}[4]{\PathsBBackRaw(#1,#2,#3,#4)}
\newcommand{\PathsForward}[3]{\textsc{Paths}_{\rightarrow}(#1,#2,#3)}
\newcommand{\PathsAndRaw}{\textsc{Paths}_{\land}}
\newcommand{\PathsAnd}[4]{\PathsAndRaw(#1,#2,#3,#4)}
\newcommand{\PathsBAndRaw}{\textsc{Paths}_{\overline{\land}}}
\newcommand{\PathsBAnd}[4]{\PathsBAndRaw(#1,#2,#3,#4)}

\newcommand{\transsf}[3]{\llbracket #1 \rrbracket(#2,#3)}
\newcommand{\transsfe}[3]{\mathsf{trans\_sf}(#1, #2, #3)}

\newcommand{\successors}[2]{\mathsf{succ}(#1,#2)}

\newcommand{\card}[1]{\ensuremath{\left|#1\right|}}

\newcommand{\FOL}{\textbf{FOL}}

\acrodef{CTL}{Computation Tree Logic}
\acrodef{FOCTL}{First-Order \ac{CTL}}
\acrodef{SMT}{Satisfaction Modulo Theories}
\acrodef{CFG}{control flow graph}
\acrodef{tCFG}{threaded control flow graph}
\acrodef{SC}{sequential consistency}
\acrodef{TSO}{total store order}
\acrodef{PSO}{partial store order}

\usepackage{url}
\urldef{\mails}\path|{mansky1, dgriffi3, egunter}@illinois.edu|

\def\titlerunning{Specifying and Executing Optimizations for Parallel Programs}
\def\authorrunning{W. Mansky, D. Griffith, \& E.L. Gunter}
\begin{document}

\title{Specifying and Executing Optimizations for Parallel Programs}
\author{William Mansky \and Dennis Griffith \and Elsa L. Gunter
\institute{Department of Computer Science, University of Illinois at Urbana-Champaign,\\
Thomas M. Siebel Center, 201 N. Goodwin, Urbana, IL 61801-2302}
\email{\mails}}
\maketitle

\begin{abstract}
Compiler optimizations, usually expressed as rewrites on program graphs, are a core part of all modern compilers. However, even production compilers have bugs, and these bugs are difficult to detect and resolve. The problem only becomes more complex when compiling parallel programs; from the choice of graph representation to the possibility of race conditions, optimization designers have a range of factors to consider that do not appear when dealing with single-threaded programs. In this paper we present {\ptrans}, a domain-specific language for formal specification of compiler transformations, and describe its executable semantics. The fundamental approach of {\ptrans} is to describe program transformations as rewrites on control flow graphs with temporal logic side conditions. The syntax of {\ptrans} allows cleaner, more comprehensible specification of program optimizations; its executable semantics allows these specifications to act as prototypes for the optimizations themselves, so that candidate optimizations can be tested and refined before going on to include them in a compiler. We demonstrate the use of {\ptrans} to state, test, and refine the specification of a redundant store elimination optimization on parallel programs.

%\keywords{optimizing compilers, parallel programming, program transformations, temporal logic, SMT solvers}
\end{abstract}

\section{Introduction}
Of the various phases of a modern compiler, optimization is generally considered to be the most complex. At the point of optimization, programs have usually been parsed and transformed into some internal representation -- most often a control flow graph, in which nodes are labeled with instructions in some intermediate language and edges represent jumps in control flow. Before generating the low-level code that actually executes on a machine, the compiler attempts to rearrange the graph to improve its time and memory performance, without changing the behavior of the program in ways that would be considered undesirable. Optimizations are often stated as complex algorithms on program code, with only informal justifications of correctness based on an intuitive understanding of program semantics. While the transformations involved may be simple, the conditions under which they are safe to apply, which often rely on extensive program analysis, are easily misstated. In practice, even widely used compilers such as GCC sometimes transform code incorrectly \cite{csmith}, and some of these bugs have been shown to result from mishandling concurrency \cite{conccsmith}. Insufficiently analyzed optimizations may result in unreliable execution of parallel code; compiler writers may even end up having to limit the scope and complexity of the optimizations they develop, in the absence of a method to demonstrate the safety of parallel optimizations.

The goal of VeriF-OPT, a Verification Framework for Optimizations and Program Transformations, is to make correct compilation more widely accessible by providing a standard approach and toolset for specifying, testing, and verifying compilers for a wide range of languages, with a particular focus on optimization and compilation of parallel programs. The core approach of the framework is a new way of looking at optimizations: as rewrites on control flow graphs with temporal logic side conditions, as first proposed by Lacey et al.~\cite{lacey}. This approach is put into practice using a domain-specific language for specifying compiler optimizations and transformations, which we call {\ptrans}. Temporal logic formulae over program graphs allow us to simply and clearly state the conditions under which an optimization should be applied. {\ptrans} has both abstract mathematical semantics, derived from its predecessor language TRANS \cite{kalvala}, and \emph{executable} semantics that can be used by compiler designers to test and refine their transformations on actual program graphs. In this paper, we present the syntax and the abstract and executable semantics of {\ptrans}, and illustrate how it can be used to rapidly prototype and test compiler optimizations. Ultimately, we hope that the approach outlined in this paper will assist compiler writers in creating complex, reliable optimizations for parallel code.

\section{The {\ptrans} Specification Language}
\label{framework}
\subsection{{\ptrans}: A Language for Parallel Program Transformation}
The basic approach of the {\ptrans} specification language is modeled after the TRANS language of Kalvala et al.~\cite{kalvala}: optimizations are specified as rewrites on program code in the form of control flow graphs, with side conditions given in \ac{CTL}.  Intuitively, the rewrite portion of an optimization expresses the transformation to be made, and the side condition characterizes the situations in which the optimization should be applied.  \ignore{Our starting point is our previous formalization of the syntax and semantics of TRANS for sequential programs \cite{transssa}.  All formalizations and proofs have been developed in the Isabelle theorem prover \cite{isabelle}, allowing us to provide strong guarantees of correctness for our verified optimizations.}
The syntax of {\ptrans} is given by the following grammar:
$$
\begin{array}{lrl}
\mathit{A}&\mathbf{::=}& \mathsf{add\_edge}(n,m,\ell)~|~\mathsf{remove\_edge}(n,m,\ell)
~|~\mathsf{split\_edge}(n,m,\ell,i) \\&|&\mathsf{replace}\;n\;\mathsf{with}\;i_1,...,i_k\\
\mathit{\varphi}\ &\mathbf{::=}& \mathsf{true}~|~p~|~\varphi \wedge \varphi~|~\neg\varphi~|~\AU{\varphi}{\varphi}~|~\EU{\varphi}{\varphi}~|~\AB{\varphi}{\varphi}~|~\EB{\varphi}{\varphi}~|~\exists x.\ \varphi\\
\mathit{T}&\mathbf{::=}& A_1,...,A_k\ \mathrm{if}\ \varphi ~|~ \mathrm{MATCH}\;\varphi\;\mathrm{IN}\;T ~|~ T\;\mathrm{THEN}\;T ~|~ T\;\square\;T ~|~ \mathrm{APPLY\_ALL}\;T
\end{array}
$$
It consists of three main syntactic categories: actions, side conditions, and transformations. The atomic \emph{actions} $A$ include $\mathsf{add\_edge}$ and $\mathsf{remove\_edge}$, which add and remove ($\ell$-labeled) edges between the specified nodes; $\mathsf{split\_edge}$, which splits an edge between two nodes, inserting a new node between them; and $\mathsf{replace}$, which replaces the instruction at a given node with a sequence of instructions, adding new nodes to contain the new instructions as necessary. \ignore{Kalvala et al. have shown that a wide variety of common program transformations can be expressed using these basic rewrites.}  The arguments to the atomic actions represent nodes and instructions in the program graph, but may contain \emph{metavariables} that are instantiated to program objects when the rewrites are applied.

The side conditions $\varphi$ of {\ptrans} are based on \ac{FOCTL}, and are built starting from a set of atomic predicates $p$. The B (``back-to'') operators are the past-time counterparts to the U (``until'') operators; for instance, $\EB{\varphi_1}{\varphi_2}$ holds when there exists some path backwards through a graph such that $\varphi_1$ holds until the previous point at which $\varphi_2$ holds. The derived ``finally'' and ``globally'' operators EF, AF, EG, AG are defined from the U operators in the usual way. 
The existential quantifier $\exists$ is used to quantify over metavariables in a formula: these metavariables may then appear in the atomic predicates of a formula, enhancing the expressive power of the side conditions.

At the top level, a transformation $T$ is built out of conditional rewrites combined with \emph{strategies}.
$A_1,...,A_k\ \mathrm{if}\ \varphi$ is the basic pairing of one or more rewrites with a temporal logic side condition.  The expression MATCH $\varphi$ IN $T$ provides an additional side condition for a set of transformations, and also allows metavariables to be bound across multiple transformations.  The THEN and $\square$ operators provide sequencing and (nondeterministic) choice respectively, and APPLY\_ALL $T$ recursively applies $T$ wherever possible until it is no longer applicable to the graph under consideration.

\subsection{Parallel Control Flow Graphs}
\label{tcfg}
The TRANS approach depends fundamentally on a notion of \ac{CFG}.  Atomic rewrites are rewrites on \acp{CFG}, and \ac{CTL} side conditions are evaluated on paths through \acp{CFG}.  Thus, we require a parallel analogue to the \ac{CFG} in order to extend the approach to parallel programs.  The particular model used here, adapted from the work of Krinke \cite{tcfg}, is the \ac{tCFG}.  In our framework, a \ac{tCFG} is simply a collection of non-intersecting \acp{CFG}, one for each thread in a program. 

\begin{definition}
A \emph{\ac{CFG}} is a tuple $(N, E, s, x, L)$ describing a labeled directed graph, where $N$ is a finite set of nodes, $E : 2^{N \times N \times T}$ is a set of labeled edges, $s, x \in N$ are the start and exit node of the graph respectively, and $L : N \rightarrow I$ is a labeling of nodes with program instructions.  The set of edge labels $T$ is provided by the target language, but must include the sequencing edge $\mathsf{seq}$. A \emph{\ac{tCFG}} is a collection of disjoint \acp{CFG}, one for each thread in the program being represented.  If $\mathcal{G}$ is a \ac{tCFG} and $t$ is a thread, we write $\mathcal{G}_t$ for the \ac{CFG} of $t$ in $\mathcal{G}$.\end{definition}
\ignore{Here we see a second parameter that must be provided by the target language: a correspondence between instruction labels and outgoing edges. For instance, in most programming languages, an assignment statement should have only one outgoing edge, indicating the next instruction to be executed; a conditional branch statement, on the other hand, should have two outgoing edges, one clearly marked as belonging to each branch. Generalizing this correspondence as a parameter allows us to reason about \acp{CFG} and \acp{tCFG} independently of any particular programming language.}

\label{preds}
The set of atomic predicates used in side conditions may depend on the target language under consideration, but some simple predicates are applicable to almost every language, and many optimizations can be specified with only language-independent predicates. These predicates include the following:
\begin{itemize}
\item $\mathsf{node}_t(n)$, which is true of a state $q$ when $q(t) = n$.
\item $\mathsf{stmt}_t(i)$, which is true of a state $q$ when the instruction at $q$ is $i$ in $\mathcal{G}_t$.
\item $\mathsf{out}_t(n', \ell)$, which is true of a state $q$ when $q(t)$ has an outgoing edge to $n'$ with label $\ell$ in $\mathcal{G}_t$.
\item $\mathsf{start}$, which is true when $q$ is at the start node of each of its component graphs, and $\mathsf{exit}$, which is true when $q$ is at the exit node for each graph.
\ignore{\item $\mathsf{is}(e_1, e_2)$, which is true when $e_1$ and $e_2$ are arithmetic expressions that can be shown to have the same value (independent of program state).}
\end{itemize}
Note that all of these predicates are static properties of \acp{tCFG} that do not depend on the semantics of the language under consideration. In general, {\ptrans} optimizations can be stated and performed independently of the semantics of the target language, so that {\ptrans} may serve as a design tool even in the absence of formal semantics for the target language.

\section{The Semantics of {\ptrans}}
\subsection{Abstract Semantics}
In this section we present the mathematical semantics of PTRANS, based on the semantics of TRANS by Kalvala et al.~\cite{kalvala}. The semantics of actions is given by a function $\llbracket A \rrbracket(\sigma, \mathcal{G})$ that takes an action, a substitution (a partial map from metavariables to program objects), and a \ac{tCFG} and returns the \ac{tCFG} that results when the action is performed. Since every action specifies at least one node and the nodes of \acp{CFG} in a \ac{tCFG} are disjoint, each action implicitly specifies at most one \ac{CFG} $\mathcal{G}_t$ on which to perform the action (if two nodes mentioned are in two different graphs, the action simply fails). Suppose we have $\mathcal{G}_t = (N_t, E_t, s_t, x_t, L_t)$; then the semantics of actions are defined as follows:
\begin{itemize}
\item $\llbracket \mathsf{add\_edge}(n, m, \ell) \rrbracket(\sigma, \mathcal{G}) = \mathcal{G}(t \mapsto (N_t, E_t \cup \{(\sigma(n), \sigma(m), \sigma(\ell))\}, s_t, x_t, L_t))$
\item $\llbracket \mathsf{remove\_edge}(n, m, \ell) \rrbracket(\sigma, \mathcal{G}) = \mathcal{G}(t \mapsto (N_t, E_t - \{(\sigma(n), \sigma(m), \sigma(\ell))\}, s_t, x_t, L_t))$
\item $\llbracket \mathsf{replace}\;n\;\mathsf{with}\; i_1, ..., i_k\rrbracket(\sigma, \mathcal{G}) = \mathcal{G}(t \mapsto (N_t \cup \{n_2, ..., n_k\}, \{\mathsf{remap\_succ}(\sigma(n), n_k, e)\ |\ e \in E\} \cup \{(n_j, n_{j+1}, \mathsf{seq})\ |\ 1 < i < k\}, s_t, x_t, L_t + (n_1 \mapsto \sigma(i_1), ..., n_k \mapsto \sigma(i_k)))$, where $n_1 = \sigma(n)$ and $n_2, ..., n_k$ are new nodes not in $\mathcal{G}$, and $\mathsf{remap\_succ}$ is defined below
\item $\llbracket \mathsf{split\_edge}(n, m, \ell, i)\rrbracket(\sigma, \mathcal{G}) = \mathcal{G}(t \mapsto (N_t \cup \{n'\}, E_t - \{(\sigma(n), \sigma(m), \sigma(\ell))\}\ \cup $\\$\{(\sigma(n), n', \sigma(\ell)), (n', \sigma(m), \mathsf{seq})\}, s_t, x_t, L_t + (n' \mapsto \sigma(i))))$, where $n'$ is a new node not in $\mathcal{G}$
\end{itemize}
In the $\mathsf{replace}$ action, we must not only introduce new $\mathsf{seq}$ edges between the added nodes, but also move the outgoing edges of the initial node $n_1$ to instead be outgoing edges of the last added node $n_k$. To do this we use the auxiliary $\mathsf{remap\_succ}$ function, defined as
$$\mathsf{remap\_succ}(n, n', (a, b, \ell)) \triangleq \text{if } a = n \text{ then } (n', b, \ell) \text{ else } (a, b, \ell)$$
The semantics of a list of actions $A_1, ..., A_k$ is the composition of the semantic functions of the individual actions, i.e., the graph resulting from applying all of the actions in sequence. 

\label{CTL}
The side conditions of {\ptrans} are given in the branching-time temporal logic \ac{FOCTL}. A \ac{CTL} formula expresses a property over a (possibly infinite) tree of \emph{states}, and at each branching point quantifies over the possible \emph{paths} forward or backward from that state (written as $\mathit{Paths}$ and $\mathit{RPaths}$ respectively; note that backward paths must always reach the start state of the graph). The formulae are made first-order by allowing variables to appear in the atomic state predicates $p$, and we can quantify over these variables with the $\exists$ operator. The semantics of an \ac{FOCTL} formula is given by a satisfaction relation of the form $\mathcal{G}, \sigma, q \models \varphi$, where $\mathcal{G}$ is a \ac{tCFG}, $\sigma$ a substitution of values for metavariables, $q$ a state (a vector of points in a \ac{tCFG}), and $\varphi$ a \ac{FOCTL} formula, defined as follows (where $\lambda_i$ denotes the $i$th element of the path $\lambda$):
\begin{itemize}
\item $\mathcal{G}, \sigma, q \models \mathsf{true}$
\item $\mathcal{G}, \sigma, q \models p$ if $\sigma(p)$ is true at $q$ in the semantics for $\sigma(p)$ provided by the target language
\item $\mathcal{G}, \sigma, q \models \varphi_1 \land \varphi_2$ if $\mathcal{G}, \sigma, q \models \varphi_1$ and $\mathcal{G}, \sigma, q \models \varphi_2$
\item $\mathcal{G}, \sigma, q \models \neg \varphi$ if $\mathcal{G}, \sigma, q \not\models \varphi$
\item $\mathcal{G}, \sigma, q \models \AU{\varphi_1}{\varphi_2}$ if $\forall \lambda\in \mathit{Paths}(\mathcal{G}, q).\ \exists i.\ \mathcal{G}, \sigma, \lambda_i \models \varphi_2 \land \forall j<i.\ \mathcal{G}, \sigma, \lambda_j \models \varphi_1$
\item $\mathcal{G}, \sigma, q \models \EU{\varphi_1}{\varphi_2}$ if $\exists \lambda\in \mathit{Paths}(\mathcal{G}, q).\ \exists i.\ \mathcal{G}, \sigma, \lambda_i \models \varphi_2 \land \forall j<i.\ \mathcal{G}, \sigma, \lambda_j \models \varphi_1$
\item $\mathcal{G}, \sigma, q \models \AB{\varphi_1}{\varphi_2}$ if $\forall \lambda\in \mathit{RPaths}(\mathcal{G}, q).\ \exists i.\ \mathcal{G}, \sigma, \lambda_i \models \varphi_2 \land \forall j<i.\ \mathcal{G}, \sigma, \lambda_j \models \varphi_1$
\item $\mathcal{G}, \sigma, q \models \EB{\varphi_1}{\varphi_2}$ if $\exists \lambda\in \mathit{RPaths}(\mathcal{G}, q).\ \exists i.\ \mathcal{G}, \sigma, \lambda_i \models \varphi_2 \land \forall j<i.\ \mathcal{G}, \sigma, \lambda_j \models \varphi_1$
\item $\mathcal{G}, \sigma, q \models \exists x.\ \varphi$ if $\exists o.\ \mathcal{G}, \sigma(x \mapsto o), q \models \varphi$
\end{itemize}
We write $\mathcal{G}, \sigma \models \varphi$ to abbreviate $\mathcal{G}, \sigma, q_0 \models \varphi$, where $q_0$ is the vector that for each \ac{CFG} in $\mathcal{G}$ gives that \ac{CFG}'s starting node.

The semantics of strategies is given by a function $\llbracket T \rrbracket(\tau, \mathcal{G})$ that takes a transformation, a substitution, and a \ac{tCFG} and returns the set of \acp{tCFG} that can be produced by the transformation. In order to give semantics to the APPLY\_ALL strategy, we must define the result of applying a transformation to a graph some finite (but unbounded) number of times:
$$\inference{ }{G \in \mathsf{apply\_some}(T, \tau, G)}~~~\inference{G' \in \llbracket T\rrbracket(\tau, G) ~~~ G'' \in \mathsf{apply\_some}(T, \tau, G')}{G'' \in \mathsf{apply\_some}(T, \tau, G)}$$
Then the semantics of strategies is defined as follows:
\begin{itemize}
\item $\llbracket A_1,...,A_k\ \mathrm{if}\ \varphi \rrbracket(\tau, \mathcal{G}) = \{\mathcal{G}'\ |\ \exists \sigma.\ \sigma|_{\mathrm{dom}(\tau)} = \tau \land \mathcal{G}, \sigma \models \varphi \land \mathcal{G}' = \llbracket A_1, ..., A_k \rrbracket(\sigma, \mathcal{G})\}$
\item $\llbracket \mathrm{MATCH}\;\varphi\;\mathrm{IN}\;T \rrbracket(\tau, \mathcal{G}) = \{\mathcal{G}'\ |\ \exists \sigma.\ \sigma|_{\mathrm{dom}(\tau)} = \tau \land \mathcal{G}, \sigma \models \varphi \land \mathcal{G}' \in \llbracket T \rrbracket(\sigma, \mathcal{G})\}$
\item $\llbracket T_1\;\mathrm{THEN}\;T_2 \rrbracket(\tau, \mathcal{G}) = \displaystyle\bigcup\limits_{\mathcal{G}' \in \llbracket T_1 \rrbracket(\tau, \mathcal{G})} \llbracket T_2 \rrbracket(\tau, \mathcal{G}')$
\item $\llbracket T_1\;\square\;T_2 \rrbracket(\tau, \mathcal{G}) = \llbracket T_1 \rrbracket(\tau, \mathcal{G}) \cup \llbracket T_2 \rrbracket(\tau, \mathcal{G})$
\item $\llbracket \mathrm{APPLY\_ALL}\;T \rrbracket(\tau, \mathcal{G}) = \mathsf{apply\_some}(\llbracket T \rrbracket, \tau, \mathcal{G}) - \{\mathcal{G}'\ |\ \exists \mathcal{G}'' \neq \mathcal{G}'.\ \mathcal{G}'' \in \llbracket T \rrbracket(\tau, \mathcal{G}')\}$
\end{itemize}
Note in particular the semantics for APPLY\_ALL, which produces the set of graphs that result from applying the transformation $T$ repeatedly in any way such that, ultimately, $T$ can no longer be applied to modify the final result.

While the semantic function for actions is straightforwardly executable (modulo suitable data structures for representing sets), the semantic function for transformations is not; it explicitly uses existential witnesses to create the (potentially infinite) set of result graphs. In particular, we frequently quantify over all substitutions that satisfy the side conditions of a transformation. In the remainder of this section, we will give a more directly executable semantics for transformations, but we must first present a method for computing the satisfying substitutions of an \ac{FOCTL} side condition.

\subsection{FOCTL Model Finding}
\label{foctl}
The model checking problem for \ac{CTL} and its variants is a well-studied problem with a
well-known efficient algorithm~\cite{clarkemc}, but considerably less
attention has been given to the related problem of \emph{model finding}. The model
finding problem in its general form is this: suppose we have an \ac{FOCTL} formula $\varphi$ built from a set
of atomic predicates, where the predicates may contain free variables. Given a transition
system $\mathcal{S}$ and an interpretation of the atomic predicates on $\mathcal{S}$, what
are the possible assignments of values to the free variables of $\varphi$ such
that $\varphi$ holds on $\mathcal{S}$? When a formula contains no free variables, model
finding is simply model checking; in the general case, it is considerably more complex. 

Following Bohn et al.~\cite{foctl}, we present an algorithm for \ac{FOCTL} model finding. The algorithm is given in a functional style and can be straightforwardly implemented in a functional programming language. 
We will present the algorithm on a single CFG; it can be extended to \acp{tCFG} via a cross
product construction. We find satisfying models \emph{symbolically}, by defining a function \textsc{Satis} that, given a formula $\varphi$ and a node $v$, constructs a non-temporal first-order 
formula characterizing the set of substitutions that make $\varphi$ true at $v$.
The following theorem states the correctness of the algorithm:

\begin{theorem}\label{mfcorrect}
Let $\mathcal{G} = (\mathcal{N},\mathcal{E},s,x,L)$ be a CFG, $v\in \mathcal{N}$ and $\varphi$ a \ac{FOCTL} formula. Then
\\$\{\sigma\ |\ \foctlmod{\mathcal{G}}{\sigma}{v}{\varphi}\}
=\{\sigma\ |\ \smtmod{\sigma}{\satis{\varphi}(v)}\}$.
\end{theorem}

The theorem is proved by induction on the lexicographic order of the number of
$\AU{\varphi_1}{\varphi_2}$-headed subformulae of $\varphi$ and the number of subformulae of
$\varphi$; we will define \textsc{Satis} and give the proof of its correctness case by case.
When the head connective is a non-temporal connective, 
\textsc{Satis} recursively translates its subformulae, leaving the connective untouched:
\begin{align*}
\satis{p(\vec{x})}(v)&=p(\vec{x})&
\satis{\varphi_1\land\varphi_2}(v)&=\satis{\varphi_1}(v)\land\satis{\varphi_2}(v)\\
\satis{\neg\varphi}(v)&=\neg\satis{\varphi}(v)&
\satis{\exists x.\ \varphi}(v)&=\exists x.\ \satis{\varphi}(v)
\end{align*}
Correctness for these cases follows directly from the inductive hypothesis.

When $\varphi=\EU{\varphi_1}{\varphi_2}$, we need to ensure that we find a suitable witness path for the until-formula
for each substitution. To do so, we define an auxilary function $\PathsBack{I}{F}{n}{v}$
that takes an invariant $I\colon\mathcal{N}\to\FOL$, a final requirement
$F\colon\mathcal{N}\to\FOL$, a path length $n$, and a node $v\in\mathcal{N}$, as follows:
\begin{align*}
\PathsBack{I}{F}{0}{v}&=F(v)\\
\PathsBack{I}{F}{n}{v}&=\PathsBack{I}{F}{n-1}{v}
                             \lor\left(I(v)
                         \land\!\!\!\bigvee_{v'\in
\successors{\mathcal{E}}{v}}\!\!\!\!\!\PathsBack{I}{F}{n-1}{v'}\right)
\end{align*} where $\successors{\mathcal{E}}{v}$ is the set of successors of $v$ in $\mathcal{E}$, i.e., $\{v'\ |\ (v, v', \ell) \in \mathcal{E}\}$.
\begin{lemma}\label{lem:pathsback}
$\PathsBack{I}{F}{n}{v}$ characterizes the set of substitutions $\sigma$ such that 
there is a path $\lambda$ from $v$ of length $k \le n$ along which $\foctlmod{\mathcal{G}}{\sigma}{\lambda_k}{F(\lambda_k)}$ and $\foctlmod{\mathcal{G}}{\sigma}{\lambda_i}{I(\lambda_i)}$ for all
$i < k$.
\end{lemma}
We can then define 
$$\satis{\EU{\varphi_1}{\varphi_2}}(v)=\PathsBack{\satis{\varphi_2}}{\satis{\varphi_1}}{\card{\mathcal{N}}}{v}$$
and finish the proof of this case by noting that, since if there is any witness 
there must be a cycle-free witness, Lemma~\ref{lem:pathsback} ensures the presence of a suitable witness.

When $\varphi=\AU{\phi_1}{\phi_2}$ we again need to look for witnesses to the until-formula, this time in a
conjunctive fashion. To do this we define the auxiliary function $\PathsAnd{I}{F}{n}{v}$ that takes an invariant
$I\colon\mathcal{N}\to\FOL$, a final requirement $F\colon\mathcal{N}\to\FOL$, a length
$n$, and a node $v\in\mathcal{N}$:
\begin{align*}
\PathsAnd{I}{F}{0}{v}&=F(v)\\
\PathsAnd{I}{F}{n}{v}&=\PathsAnd{I}{F}{n-1}{v}
                     \lor\left(I(v)\land\bigwedge_{v'\in\successors{\mathcal{E}}{V}}\PathsAnd{I}{F}{n - 1}{v'}\right)
\end{align*}
\begin{lemma}\label{lem:pathsand}
$\PathsAnd{I}{F}{n}{v}$ characterizes the set of substitutions $\sigma$ such that
for every path $\lambda$ from $v$ that is of length $n$ or reaches the exit node in fewer than $n$ steps,
there is some $i$ where $\foctlmod{\mathcal{G}}{\sigma}{\lambda_i}{F(\lambda_i)}$ and $\foctlmod{\mathcal{G}}{\sigma}{\lambda_i}{I(\lambda_j)}$ for all $j < i$.
\end{lemma}
This correctness lemma is more difficult to state, since the conjunctive search and the
presence of a sink (the exit node) means we must carefully handle paths with length less than
$|\mathcal{N}|$. Unfortunately, while this gives us an ability to say that we can ``always''
find a witness for our until-formula, this function by itself still allows for infinite paths that
never reach their satisfying witness. We need one more auxilary function to
help us avoid these infinite counterexamples, defined below.
\begin{align*}
\PathsForward{I}{0}{v}&=\textsc{True}\\
\PathsForward{I}{n}{v}&=\bigvee_{v'\in\successors{\mathcal{E}}{v}}\left(I(v')
                               \land\PathsForward{I}{n-1}{v'}\right)
\end{align*}

\begin{lemma}\label{lem:pathsforward}
$\PathsForward{F}{n}{v}$ characterizes the set of substitutions $\sigma$ such that there is a path $\lambda$ from $v$ of length $n$ along which $\foctlmod{\mathcal{G}}{\sigma}{\lambda_i}{I(\lambda_i)}$ for every $i$.
\end{lemma}
We can then define \[
\satis{\AU{\varphi_1}{\varphi_2}}(v)=\begin{array}[t]{l}
           \neg\PathsForward{\satis{\varphi_1\land\neg\varphi_2}}{\card{\mathcal{N}}+1}{v}\\
           \land\ \PathsAnd{\satis{\varphi_2}}{\satis{\varphi_1}}{\card{\mathcal{N}}}{v}
           \end{array}\]
We use our two auxilary functions to ensure that every path from $v$ has a suitable
witness for $\varphi_2$ and that, by the pigeonhole principle, it has no paths along which $\varphi_1$ holds and $\varphi_2$ is never reached. This case demonstrates why simple induction on the size of $\varphi$ is not sufficient to prove correctness; $\satis{\AU{\varphi_1}{\varphi_2}}(v)$ makes recursive calls on strictly larger formulae than $\AU{\varphi_1}{\varphi_2}$, but those subformulae have fewer AU-connectives.

In handling the past-time connectives we have a slight advantage: since all paths backward must eventually reach the start node of the graph, we can ignore the possiblity of infinite paths. We make an additional simplifying
assumption that all nodes in the graph are reachable from the start node (unreachable nodes can safely be discarded). This allows us to handle the
backwards cases with the duals of $\PathsBackRaw$ and $\PathsAndRaw$ formed by following
predecessors instead of successors, denoted $\PathsBBackRaw$ and $\PathsBAndRaw$
respectively.
\begin{align*}
\satis{\EB{\varphi_1}{\varphi_2}}(v)&=\PathsBBack{\satis{\varphi_2}}{\satis{\varphi_1}}{\mathcal{N}}{v}\\
\satis{\AB{\varphi_1}{\varphi_2}}(v)&=\PathsBAnd{\satis{\varphi_2}}{\satis{\varphi_1}}{\mathcal{N}}{v}
\end{align*}
This completes the definition of the \textsc{Satis} function. The running time of the algorithm as written is
$O\left(\card{\mathcal{N}}^{\card{\mathcal{N}}\card{\phi}}\right)$, but we can do better using dynamic programming. We
begin with a table of $O(\card{\phi}\card{\mathcal{N}})$ entries for $\textsc{Satis}$ and
tables of size $O(\card{\mathcal{N}})$ for the each of the path-searching functions, representing the results for each possible input given the size of $\mathcal{G}$ and the subformulae of $\varphi$. Filling each of the path tables
(assuming their arguments are already evaluated) takes $O(\card{\mathcal{N}}^2)$ time, and the tables can
be reused to answer their queries for all vertices. Thus, the amortized running time for filling an entry in the table for \textsc{Satis} is $O(\card{\mathcal{N}})$, and the overall reduction runs in $O(\card{\phi}\card{\mathcal{N}}^2)$ time.

Once we have the characteristic formula provided by \textsc{Satis}, if the basic language of atomic predicates is amenable to SMT solving, we can use a solver to compute the concrete set of satisfying models.

\subsection{Executable Semantics for Strategies}
\label{exec}

Given the model finding algorithm described above, we can define a function $\mathsf{get\_models}(\tau, \mathcal{G}, \varphi)$ that computes the satisfying models of $\varphi$ by generating a first-order formula that represents the set of substitutions that satisfy $\varphi$, conjoining it with a formula describing the already-known substitution $\tau$, and then using an SMT solver to find all satisfying models of that formula. Theorem~\ref{mfcorrect} then assures us that $\mathsf{get\_models}(\tau, \mathcal{G}, \varphi) = \{\sigma\ |\ \mathcal{G}; \sigma \vDash \varphi \land \sigma|_{\mathrm{dom}(\tau)} = \tau\}$, and so $\mathsf{get\_models}$ serves as an executable method for finding satisfying models of {\ptrans} side conditions.
Using $\mathsf{get\_models}$, we can write an executable function $\mathsf{trans\_sf}$ that finds the semantics of a transformation, defined as follows (recall that the abstract semantic function for actions is already executable):
\begin{itemize}[itemindent=-2.05in, leftmargin=2.455in]
\item $\transsfe{A_1,...,A_k\ \mathrm{if}\ \varphi}{\tau}{\mathcal{G}} = \{\text{for each }\sigma\text{ in }\mathsf{get\_models}(\tau, \mathcal{G}, \varphi), \llbracket A_1, ..., A_k \rrbracket(\sigma, \mathcal{G})\}$
\item $\transsfe{\mathrm{MATCH}\;\varphi\;\mathrm{IN}\;T}{\tau}{\mathcal{G}} = \{\text{for each }\sigma\text{ in }\mathsf{get\_models}(\tau, \mathcal{G}, \varphi), \transsfe{T}{\sigma}{\mathcal{G}}\}$
\item $\transsfe{T_1\;\mathrm{THEN}\;T_2}{\tau}{\mathcal{G}} = \displaystyle\bigcup\limits_{\mathcal{G}' \in \transsfe{T_1}{\tau}{\mathcal{G}}} \transsfe{T_2}{\tau}{\mathcal{G}'}$
\item $\transsfe{T_1\;\square\;T_2}{\tau}{\mathcal{G}} = \transsfe{T_1}{\tau}{\mathcal{G}} \cup \transsfe{T_2}{\tau}{\mathcal{G}}$
\item $\transsfe{\mathrm{APPLY\_ALL}\;T}{\tau}{\mathcal{G}} = \text{let }R = \transsfe{T}{\tau}{\mathcal{G}}\text{ in} \\
\text{if }R = \{\mathcal{G}\} \text{ then } R \text{ else }\displaystyle\bigcup\limits_{\mathcal{G}' \in R} \transsfe{\mathrm{APPLY\_ALL}\;T}{\tau}{\mathcal{G}'}$
\end{itemize}

In order to define $\mathsf{trans\_sf}$ as an executable function, we must give up on faithfully representing infinite results. In particular, our algorithm's treatment of the $\mathrm{APPLY\_ALL}$ strategy does not have exactly the same semantics as $\llbracket\mathrm{APPLY\_ALL}\rrbracket$. In the abstract semantics, we used $\mathsf{apply\_some}$ to describe the set of results produced by applying a transformation $T$ some finite number of times, and subtracted the result graphs that could still be further transformed; if $T$ could transform a graph $\mathcal{G}$ indefinitely, the infinite sequence of rewrites would contribute nothing to $\transsf{\mathrm{APPLY\_ALL}\;T}{\tau}{\mathcal{G}}$. The $\mathsf{trans\_sf}$ function, on the other hand, attempts to apply $T$ to $\mathcal{G}$ indefinitely, and so will never terminate. However, in all finite cases it can be shown that $\transsfe{T}{\tau}{\mathcal{G}} = \transsf{T}{\tau}{\mathcal{G}}$, and so $\mathsf{trans\_sf}$ is a viable executable semantics for {\ptrans} transformations.

This gives us an algorithm for computing the result graphs for a given transformation, which can be implemented in a functional language (we have chosen F\# for its Z3 integration). As long as a transformation expressed in {\ptrans} does not require infinite computations, we can run it on a target graph and obtain all of its outputs. In the following section, we will demonstrate the use of these semantics to define, test, and refine a sample optimization.

\section{Designing and Prototyping Optimizations with {\ptrans}}
\subsection{A Sample Target Language: MiniLLVM}
\label{language}
In this section, we will develop an optimization in {\ptrans} and show how its executable semantics can be of use in the design process. We will begin by defining a target language: MiniLLVM, a simplification of the LLVM intermediate language \cite{llvm}. The syntax of MiniLLVM is as follows:
$$\mathit{expr} ::= \mathtt{\%}\mathrm{x}~|~\mathtt{@}\mathrm{x}~|~\mathrm{c} ~~~~~~~~~~~~~ \mathit{type} ::= \mathtt{int}~|~\mathit{type}^\mathtt{*}$$
\vspace{-.2in}
$$\begin{array}{rl}
\mathit{instr} ::=& \mathtt{\%}\mathrm{x} = \mathrm{op}\ \mathit{type}\ \mathit{expr}\mathtt{,}\ \mathit{expr}\ | \%\mathrm{x} = \mathtt{icmp}\ \mathrm{cmp}\ \mathit{type}\ \mathit{expr}\mathtt{,}\ \mathit{expr}\ |\ \mathtt{br}\ \mathtt{i1}\ \mathit{expr}~|~\mathtt{br}~|~\\&\%\mathrm{x} = \mathtt{call}\ \mathit{type}\ (\mathit{expr}\mathtt{,}\ ...\mathtt{,}\ \mathit{expr})~|~\mathtt{return}\ \mathit{expr}\ |\ \%\mathrm{x} = \mathtt{alloca}\ \mathit{type}~|~\\&\mathtt{\%}\mathrm{x} = \mathtt{load}\ \mathit{type}^\mathtt{*}\ \mathit{expr}~|~\mathtt{store}\ \mathit{type}\ \mathit{expr}\mathtt{,}\ \mathit{type}^\mathtt{*}\ \mathit{expr}~|~\mathtt{is\_pointer}\ \mathit{expr}\ignore{\mathtt{\%}\mathrm{x} = \mathtt{cmpxchg}\ \mathit{type}^\mathtt{*}\ \mathit{expr}\mathtt{,} \mathit{type}\ \mathit{expr}\mathtt{,} \mathit{type}\ \mathit{expr}~|~}
\end{array}$$
MiniLLVM expressions are either local variables ($\mathtt{\%}\mathrm{x}$), global variables ($\mathtt{@}\mathrm{x}$), or constants. Instructions include arithmetic operations (where op is an arithmetic operator), comparison operations (where cmp is a comparison operator), conditional and unconditional branches, function calls and returns, memory allocation, loads from and stores to memory, and $\mathtt{is\_pointer}$, which checks whether a given expression is pointer-valued (for use in loads and stores). (Note that the $^\mathtt{*}$'s indicate not repetition but pointer types.) Because the targets of control-flow instructions are encoded in the edges of the \ac{CFG}, the label arguments to $\mathtt{br}$ instructions and function names in $\mathtt{call}$ instructions are omitted. Although each $\mathtt{alloca}$ instruction is executed by a single thread, the memory allocated can be exposed to other threads by storing its location in a global variable or fixed memory location. 

For the purposes of this example, we will assume that MiniLLVM has a straightforward interleaved semantics, with a sequential consistency memory model: i.e., in each step one thread in the program executes, and any memory operations immediately update the shared memory and are visible to all other threads. More relaxed memory models, such as total or partial store ordering, are important to consider when designing compiler optimizations on parallel programs, and any operational memory model can be integrated into the semantics of MiniLLVM (and thus the optimization testing process) with little difficulty.

\subsection{Writing {\ptrans} Optimizations}
\label{opts}
\begin{figure}[htbp]
\centering
\includegraphics[scale=0.4]{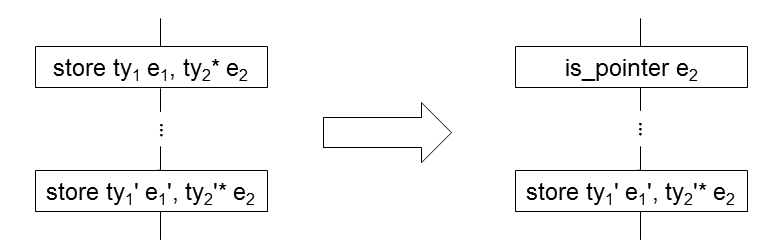}
\caption{Redundant Store Elimination}
\label{pattern}
\end{figure}

Now that we have a target language, we can begin to define {\ptrans} transformations on MiniLLVM. Our case study will be a \emph{redundant store elimination} optimization (RSE), which removes stores that may be overwritten before they are used, as shown in Figure~\ref{pattern}. Note that the redundant store is replaced by an $\mathtt{is\_pointer}$ instruction, rather than being eliminated entirely, to ensure that crashes are not delayed in bad executions in which $e_2$ is not a pointer-valued expression.

The rewrite involved is simple: replace the instruction at the chosen node with the $\mathtt{is\_pointer}$ instruction. The side condition should require that there is a node $n$ containing the store to be eliminated, and that along all paths forward from $n$ another store occurs that makes $n$ redundant. To make the optimization safe, we must also require some property to hold on the instructions between $n$ and the following store. For instance, if the value of $e_2$ is changed before the next store to $e_2$, removing the store at $n$ would change the behavior of the program. Let $\varphi$ be some restriction on the types of instructions that can appear between $n$ and the stores that make it redundant; then the {\ptrans} specification of RSE can be written as:
$$\begin{array}{ll}\mathit{RSE}(\varphi) \triangleq &\mathsf{replace}\ n\ \mathsf{with}\ \mathtt{is\_pointer}\ e_2\ \mathrm{if}\\ &\mathrm{EF}\ \mathsf{node}_t(n) \land \mathsf{stmt}_t(\mathtt{store}\ \mathit{ty}_1\ e_1, \mathit{ty}_2^{*}\ e_2)\ \land \\ &\AU{\varphi}{(\neg\mathsf{node}_t(n) \land \mathsf{stmt}_t(\mathtt{store}\ \mathit{ty}_1'\ e_1', \mathit{ty}_2'^{*}\ e_2))}\end{array}$$

To finish this definition, we must find a suitable value for $\varphi$.
The most precise form of RSE would involve using alias analysis to determine whether memory operations may, must, or cannot refer to the location indicated by $e_2$ at $n$. For the purposes of our example, we will instead give a conservative approximation of the necessary condition, one that guarantees the safety of the transformation but may miss some redundant stores. First, we will need to require that the value of $e_2$ is not changed, so that we know that successive stores to $e_2$ do indeed overwrite the store at $n$; we can do this through the use of a defined $\mathsf{def}$ predicate describing all the instructions that might redefine a variable (recall that MiniLLVM expressions are either constants, or local or global variables). 
 We will also need to place some restriction on the kinds of memory operations that can be performed between $n$ and a following store; after all, if the value stored to $e_2$ is used in any way before being overwritten, the store is not redundant. In the absence of alias analysis, we must assume that any reference to a memory location could overlap with $e_2$, so our condition must rule out any $\mathtt{load}$ instructions between $n$ and a following store. We can define a predicate $\mathsf{not\_loads}$ such that $\mathsf{not\_loads}_t(e)$ is true when the instruction in $t$ is not a load from $e$, and then write the remainder of our side condition as $\varphi_1 \triangleq \neg\mathsf{def}_t(e_2) \land \forall e.\ \mathsf{not\_loads}_t(e)$.

\begin{figure}[htb]
\begin{subfigure}[b]{0.35\textwidth}
\centering
\includegraphics[scale=0.4]{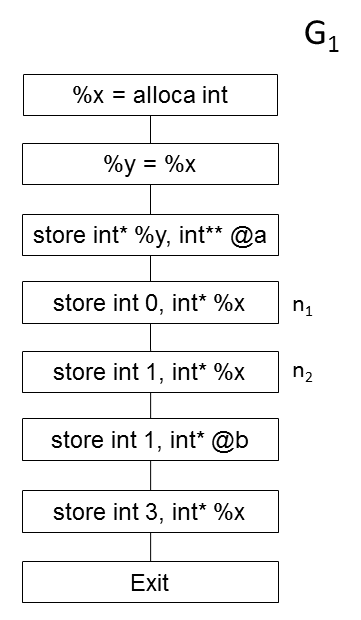}\ \ %
\caption{A graph with two redundant stores}
\label{ex1}
\end{subfigure}
\quad
\begin{subfigure}[b]{0.5\textwidth}
\includegraphics[scale=0.4]{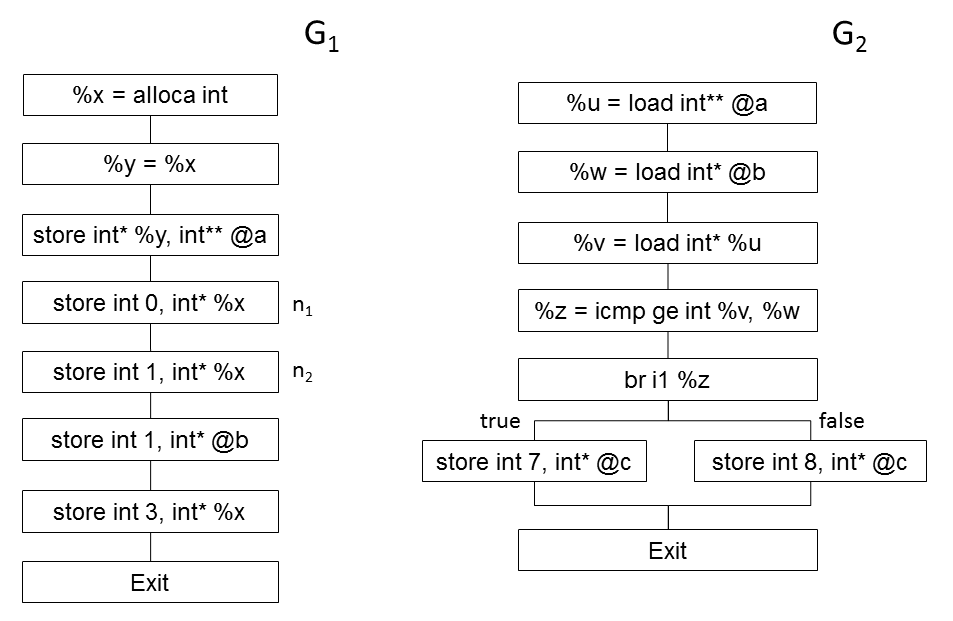}
\caption{A \ac{tCFG} with redundant stores?\qquad}
\label{ex2}
\end{subfigure}
\caption{An RSE example}
\end{figure} 

Using our executable semantics, we can run $\mathit{RSE}(\varphi_1)$ on a range of example \acp{CFG}, such as the graph $\mathrm{G_1}$ shown in Figure~\ref{ex1}. The program in $\mathrm{G_1}$ initializes a local pointer $\mathtt{\%x}$, creates an alias to it in $\mathtt{\%y}$ and publicizes its location in the global variable $\mathtt{@a}$, and then performs a series of stores to shared memory.
The $\mathsf{trans\_sf}$ function will give us two possible results for $\mathit{RSE}(\varphi_1)$ on $\mathrm{G_1}$, one in which each of $\mathrm{n_1}$ and $\mathrm{n_2}$ is replaced by an $\mathtt{is\_pointer}$ instruction (we could also use $\mathrm{APPLY\_ALL}\ \mathit{RSE}(\varphi_1)$ to apply the transformation repeatedly, replacing both $\mathrm{n_1}$ and $\mathrm{n_2}$). Furthermore, running each of the transformed programs shows that they produce the same results as the original program: 0 at the location of $\mathtt{\%x}$, the value of $\mathtt{\%x}$ at $\mathtt{@a}$, and 1 at $\mathtt{@b}$. Thus far, $\varphi_1$ appears to be a sufficient condition to ensure the correctness of RSE, and this condition is indeed sufficient for single-threaded programs.

However, when we expand our aims to parallel programs, a potential error becomes apparent. Consider the \ac{tCFG} in Figure~\ref{ex2}.
Although the program is not well synchronized, we can see that the $\mathsf{false}$ branch in $\mathrm{G_2}$ will never be taken, since if we successfully read the value at $\mathtt{@b}$ into $\mathtt{\%w}$, a value greater than or equal to 1 will have already been stored to $\mathtt{\%x}$. However, if the store at $\mathrm{n_2}$ is removed, then we may reach a state in which $\mathtt{\%w}$ is 1 and $\mathtt{\%v}$ is 0, allowing the value 8 to be stored in $\mathtt{@c}$. This means that $\mathit{RSE}(\varphi_1)$ will introduce new observable behaviors in the \ac{tCFG}: in the original graph the final value of the global variable $\mathtt{@c}$ is always 7, but in the transformed graph it may be 8. Correct optimizations may rule out some executions (for instance, by optimizing away an outcome of a race condition), but they should never introduce new behavior. Thus, this test case shows that we need to tighten the condition on our RSE optimization to make it safe on parallel programs.

The simplest refinement is to disallow any changes to shared memory between a store to be removed and its following stores. In the example above, if the store to $\mathtt{@b}$ in $\mathrm{G_1}$ did not exist, then it would be impossible for $\mathrm{G_2}$ to distinguish between the case in which the store at $\mathrm{n_2}$ was removed and the one in which it had already been overwritten by the final store to $\mathtt{\%x}$. Since we have already ruled out $\mathtt{load}$ instructions, we need only prohibit $\mathtt{store}$ instructions as well; the appropriate side condition in {\ptrans} can be written as $\varphi_2 \triangleq \neg\mathsf{def}_t(e_2) \land ((\forall e.\ \mathsf{not\_loads}_t(e) \land \mathsf{not\_stores}_t(e)) \lor \mathsf{node}_t(n))$, where we add a special case to allow for the possibility of looping back through $n$ before reaching the following store. Running $\mathsf{trans\_sf}$ on $\mathit{RSE}(\varphi_2)$ will then remove the store at $\mathrm{n_1}$, but leave $\mathrm{n_2}$ untouched. We can run the resulting program and see that, as desired, the transformed program will never produce a value of 8 in $\mathtt{@c}$. Through the process of iterated testing and refinement, we have produced an apparently correct form of the RSE optimization on parallel programs -- although, if later tests show $\varphi_2$ to be insufficient to ensure correctness, we can repeat the process and devise a still stronger condition.

\section{Implementation}
\label{impl}
We have implemented the executable semantics of {\ptrans} described above in F\#~\cite{fsharp}, taking advantage of its integration with the Z3 SMT solver \cite{z3}. The semantic functions for actions and strategies in Section~\ref{exec} can be straightforwardly translated into F\# code. We use the algorithm of Section~\ref{foctl} to reduce side conditions to first-order formulae that can be passed to Z3, and make repeated calls to Z3 to get all the satisfying models, in each iteration adding a condition that rules out the previous model. We memoize the \textsc{Satis} function with a standard lookup table in order to achieve the desired running time. The examples of Section~\ref{opts} complete in between 1 and 4 seconds, with the majority of the running time devoted to constructing the SMT queries; we believe that further optimization of the condition-generation process will allow the semantics to scale to more extensive program graphs.

\section{Related Work}
Our work builds on the TRANS approach of expressing optimizations as rewrites on control flow graphs with temporal logic side conditions due to Lacey et al. \cite{lacey} and Kalvala et al. \cite{kalvala}. The most closely related tool is Cobalt \cite{cobalt}, a system for specifying optimizations in a TRANS-like language. Cobalt optimizations are both executable and automatically verified, though it provides no support for iterative refinement of possibly incorrect specifications. Automation also comes at the cost of expressiveness: Cobalt is limited to a much smaller set of CTL side conditions than TRANS or {\ptrans}, and thus can express a smaller range of optimizations. To the best of our knowledge, neither Cobalt nor any other work stemming from the TRANS approach has yet addressed the question of parallelism.

He and Bowen \cite{logiccomp} have also developed a language for specifying and prototyping compiler transformations, focusing particularly on the code generation phase of compilation. Their language consists of if-expressions analogous to our strategy-free transformations, and is implemented as a set of Horn clauses in Prolog. Rather than giving operational semantics for a real-world target language, they model programs directly as sequences of modifications to the machine state. Because they deal primarily with language-to-language translation, their transformations are not innately composable, and they deal largely with local peephole optimizations rather than those involving dataflow analysis.

CompCert \cite{compcert}, the definitive example of a proof of compiler correctness, includes a Coq-based framework for specifying and verifying compiler optimizations; executable semantics are obtained by extracting code from the Coq definition, guaranteeing its correctness. Their specifications follow the traditional algorithmic approach to dataflow analysis, with the conditions under which an optimization should be applied expressed as a set of transfer functions for dataflow equations, and must be written as instances of Coq functors rather than as a separate domain-specific language. The ongoing CompCertTSO project \cite{conccompcert} seeks to add support for concurrency to CompCert, and has involved the specification and verification of a small number of concurrency-specific optimizations, as well as a range of sequential optimizations that can be lifted as-is to the concurrent case. CompCertTSO also verifies each translation in the chain from high-level source language to low-level machine code, while the work presented here is limited to transformations within a single target language. We believe that the most significant advantage of our approach is its language- and memory-model independence; we make no particular assumptions that restrict us to MiniLLVM or a particular treatment of concurrent memory models.

\section{Conclusion and Future Work}
In this paper, we show the use of the {\ptrans} specification language in designing and prototyping compiler optimizations for parallel programs in terms of graph transformations. By expressing optimizations as rewrites on control flow graphs with temporal logic side conditions, {\ptrans} allows for a more direct expression of the logic behind transformations. The mathematical semantics of {\ptrans} are accompanied by an executable semantics, allowing us to run {\ptrans} specifications directly on program graphs. The executable semantics relies on an algorithm for finding satisfying models of first-order CTL formulae on a graph, which in combination with an SMT solver can efficiently find all possible locations at which a transformation applies. {\ptrans}, with its combination of abstract and executable semantics, lays the groundwork for a unified platform for specifying, testing, and verifying optimizations. 

While we have implemented the executable semantics of {\ptrans} in F\# with Z3 integration, we are also interested in developing it in the K Framework \cite{koverview} for programming language specification. A K implementation of {\ptrans} could take advantage of built-in state-space search functionality, as well as the wide range of languages that have been given formal semantics in K, including C, OCaml, and a fuller version of LLVM. We also intend to move forward with the formal verification of optimizations specified in {\ptrans} in the Isabelle theorem prover \cite{isabelle}, and ultimately hope to link our executable semantics with our abstract semantics through a formal soundness proof.

\paragraph{Acknowledgements}
\sloppy
This material is based upon work supported in part by NSF Grant CCF 13-18191. Any opinions, findings, and conclusions or recommendations expressed in this material are those of the authors and do not necessarily reflect the views of the NSF.
\fussy
\bibliographystyle{eptcs}
\bibliography{sources}

\end{document}